\documentclass[hidelinks]{IEEEtran}

\usepackage[style=ieee, doi=true]{biblatex}
\usepackage{verbatim}
\usepackage{graphicx} 
\usepackage{todonotes}
\usepackage{hyperref}
\usepackage{amsmath}

\DeclareFieldFormat{doi}{\href{https://doi.org/#1}{https://doi.org/#1}}
\renewbibmacro*{doi+eprint+url}{%
  \printfield{doi}%
  \newunit\newblock%
  \iffieldundef{doi}{\printfield{url}}{}%
}
\begin{document}
 \author{
 \IEEEauthorblockN{Yannic Talavera},
 \IEEEauthorblockA{Hamm am Rhein, Germany,
  Email: yannic.talavera@gmail.com}
 \and
 \IEEEauthorblockN{Bernd Ulmann},
 \IEEEauthorblockA{anabrid GmbH /
  FOM University of Applied Sciences,
  Frankfurt/Main, Germany,
  Email: ulmann@anabrid.com\\}
 }
\title{Brain Organoid Computing -- an Overview}



\maketitle
\pagebreak
\begin{abstract}The aim of this paper is to give an overview of \emph{brain organoid computing}, its characteristics, challenges, as well as possible advantages for future applications in the field of artificial intelligence. An important part is the extensive bibliography covering all relevant aspects and questions on this topic. Brain organoids—three-dimensional in vitro neural structures derived from human stem cells—have recently garnered attention not only in medical research but also as potential substrates for unconventional computing. Their biological nature allows them to exhibit learning behavior, plasticity, and parallel information processing, making them fundamentally different from traditional silicon-based systems. This opens up new perspectives on how intelligent systems might be designed in the future. Using brain organoids for computing presents a possible pathway towards more adaptive, energy-efficient, and biologically inspired forms of AI. However, challenges persist, particularly regarding lifespan, interfacing, reproducibility, and ethical concerns regarding the use of human-derived tissue. 
This paper aims to provide a foundational understanding for researchers exploring the convergence of human biology and computation.
\end{abstract}
\section{Introduction}\label{sec1}
Recent advances in biomedical engineering practices have led to increased research on the use of in vitro brain cultures known as \emph{brain organoids} to test responses to stimuli and investigate neural structures \cite{mulder_beginners_2023}. Despite the mainly medical focus \cite{koo_past_2019} of this research, a novel field of using these neural structures for computational tasks was created \cite{smirnova_organoid_2023,jordan_open_2024}.

Research on using biological tissue for computational purposes has so far largely been focused on using biological components to create circuits that mimic current silicon-based computers. However, this method of biocomputing forgoes the use of the full potential of cells in this context \cite{grozinger_pathways_2019}. Yet, novel research in the field of leveraging brain organoids for computing has enabled early applications, such as FinalSpark’s \emph{Neuroplatform}, that relies on human brain organoids and facilitates remote access for universities and institutions to experiment with their organic processors \cite{jordan_open_2024}. The use of brain organoids and more primitive neuronal cultures for computing-related tasks is a field with a comparatively short history. Since the advent of \emph{induced pluripotent stem cells} (\emph{iPSCs}), research on such use cases has ranged from neuronal cultures performing basic image processing \cite{ruaro_toward_2005}, moving avatars in virtual environments \cite{bakkum_spatio-temporal_2008}, to blind source separation \cite{isomura_cultured_2015}.

Following the creation of the first brain organoids, subsequent research on the use of in vitro neurons for computing has generated quite some interest \cite{smirnova_organoid_2023}. Despite the still speculative capabilities of brain organoids, researchers have begun to define and investigate how these can be further exploited for computing. Accordingly, current research focuses on developing platforms and architectures to enable novel utilizations of brain organoids in computing contexts, as well as improving the very organoids forming the substrate of this technology \cite{smirnova_organoid_2023,jordan_open_2024,johns_hopkins_university_first_2024}.

The organoids used in experiments are generated using iPSCs, placed in a suitable medium, and connected to a \emph{multielectrode array} (\emph{MEA}) \cite{govindan_mass_2021}. The MEAs allow researchers to read out electrical signals from the organoids as well as to electrically stimulate them in order to influence their behavior \cite{bakkum_spatio-temporal_2008,kagan_vitro_2022}. Using these methods researchers have been able to modify the organoids' activity by providing feedback by means of electrical signals \cite{bakkum_spatio-temporal_2008,jordan_open_2024}. Thus, such brain organoids are addressable and respond to stimuli, opening the possibility of influencing their activity and steering their processing capabilities for computational tasks.
\section{Opportunities and Challenges}
With traditional silicon processors approaching physical limits with respect to integration density, clock frequencies, etc., processors are iteratively providing less performance gain per generation \cite{leiserson_theres_2020} than in previous decades. This means that the main process driving hardware performance improvements is becoming harder to maintain and may soon stop delivering the gains that have been crucial to the advancement of technology today. Furthermore, increasing integration densities, despite the gradual deceleration in their progress, leads to quicker degradation of logic gates, lowering overall performance over time, which also limits further advances in this direction \cite{cao_future_2023}. 

Additionally, modern silicon-based processing units are impractical for research in fields like AI due to their high power requirements. Current top-of-the line machine learning models already require electric power at a Megawatt scale \cite{smirnova_organoid_2023}. Estimates show that data centers used for big data and AI applications accounted for 460 terawatt-hours in 2022 alone \cite{iea_electricity_2024}, which is even higher than France's energy consumption in 2022 \cite{eia_electricity_2024}. 
In addition, today's predominantly \textsc{von Neumann} processors are limited with respect to the achievable amount of parallelism \cite{schuman_opportunities_2022}. They also struggle to deal with decision-making based on complex and uncertain datasets \cite{smirnova_organoid_2023}, both being advantageous for AI tasks \cite{basu_organoid_2024}. 
Therefore, current processors use vast amounts of electricity for tasks they are not optimized for, while traditional methods of improving processing power are reaching fundamental limits.
Consequently, researchers have been exploring alternative methods of computing such as using organic tissue that has been shown to exhibit intelligent behavior processing information, stimuli and contexts, mimicking real biological systems.
\subsection{Potential Benefits}

\subsubsection{Adaptivity}
Brain organoids can adapt to stimuli and given tasks, successively increasing their competence for the current task and have been shown to do so within a range of activities. This ability to modify and strengthen behaviors and neural structures in response to experiences and stimuli is known as plasticity \cite{jeon_distinctive_2023}.
In vitro cultures of neurons have, for example, learned to play \emph{Pong} \cite{kagan_vitro_2022}, learned to conduct speech recognition and simple computations \cite{cai_brain_2023, }. In said examples the systems continuously improved their performance. Experiments like these showcase the capability of in vitro neural structures to plastically learn from experiences and change their behavior to optimize performance.

This is in contrast with traditional computing efforts, where the underlying hardware architecture is rigid and only software and its parametrization can be changed dynamically. The adaptability of these cultures enables an architectural plasticity that can complement the adaptability of the software, enabling flexible and efficient computing pathways \cite{basu_organoid_2024}. 

This advantage not only applies to novel tasks and problems, but could also help with traditional machine learning tasks by minimizing learning epochs \cite{cai_brain_2023,jeon_distinctive_2023}. \emph{Biological neural network}s (\emph{BNN}s) and their inherent plasticity could therefore enable systems to learn as biological brains do and progressively improve their performance in given tasks, improving efficiency. Furthermore, their plasticity allows them to process inputs predictively. This stems from the deeply interconnected nature of BNNs and their evolutionarily advantageous attempts to minimize surprises to the system by anticipating and extrapolating future activity based on past inputs \cite{luczak_neurons_2022,jeon_distinctive_2023}.

Meanwhile, contemporary \emph{artificial neural network}s (\emph{ANN}s) predominantly rely on interpolation within the bounds of their training data, making them less adaptable to unpredictable environments and data outside of their training datasets \cite{bonnasse-gahot_interpolation_2022,taheri_generalization_2024}. Experiments comparing BNNs and ANNs highlight this, with BNNs significantly outperforming ANNs in pattern recognition in randomized shapes \cite{fleuret_comparing_2011}.

This basic capability is further improved by their inherent plasticity. Accordingly, BNNs can adapt and respond quickly to new stimuli while the network improves over time to effectively deal with a wider range of stimuli \cite{basu_organoid_2024, stoop_note_2021,luczak_neurons_2022}. These capabilities are well demonstrated by BNNs adapting to perform blind source separation \cite{isomura_cultured_2015}, as well as predicting auditory stimuli \cite{luczak_neurons_2022}. In both cases the networks were exposed to training sets for short periods of time. 
\subsubsection{Parallelism}
One of the main disadvantages of traditional \textsc{von Neumann} or Harvard \emph{central processing unit} (\emph{CPU}) architectures is the inherent sequential processing of instructions. This is in stark contrast to BNNs which exhibit full parallelism due to the absence of any central control logic, memory, etc. \cite{schuman_opportunities_2022}. This characteristic has the potential to solve multiple tasks in parallel \cite{ruaro_toward_2005,basu_organoid_2024}, thus decreasing processing times and therefore creating a vastly more efficient system than one that is based solely on sequential (digital) processing.

A further advantage of brain organoid computing lies in its unconventional signaling and information communication method. Neurons emit electrical signals when firing, so-called \emph{spikes} or \emph{action potentials} \cite{sharf_functional_2022} in contrast to bits used in conventional computers.\footnote{It should be noted that such spiking behavior occurs in most but not all biological neurons.} 
Spikes transmit more information than just an on / off state as they encode information in their amplitude, waveform, as well as their duration, and the overall spiking patterns \cite{sharf_functional_2022}. Spikes are not only triggered by stimuli but can also occur spontaneously \cite{sharf_functional_2022}. 

These multidimensional signals allow for a comparatively high amount of information encoded in transmitted signals \cite{jeon_distinctive_2023}. Additionally, spikes are processed by neurons on an event-driven basis, meaning that neurons perform work mainly when spikes are present in the network \cite{schuman_opportunities_2022}.
\subsubsection{Efficiency}
Accordingly, this results in a highly energy efficient processing \cite{liu_unleashing_2024}. Using brain organoids for computational tasks could eventually result in massive efficiency gains for complex information processing systems. Given the current and projected energy requirements of modern data centers, this is a key benefit.

Hewlett Packard Enterprise’s \emph{Frontier} supercomputer requires around 24.6 megawatts of electricity to achieve an average performance of 1.353 EFLOPS ($10^{18}$ floating point operations per second) \cite{TOP500_2024}, which, although not easily, can be compared to the raw computational power of a human brain. Several sources estimate a human brain capable of performing between $10^{15}$\cite{kurzweil_singularity_2005} and $10^{18}$\cite{madhavan_brain-inspired_2023} FLOPS using roughly 20 Watts \cite{smirnova_organoid_2023}. This makes a brain more power efficient by a factor ranging from $10^3$ to $10^{6}$ -- a remarkable disparity. Biological brains are fundamentally more energy efficient than any modern digital system.

Training a large language model like OpenAI’s \emph{GPT-3} is estimated to require around 1,300 MWh of electricity \cite{luccioni_estimating_2022}, which is equivalent to the electricity consumption of more than 800 households in the European Union during one year \cite{eurostat_electricity_2024}. 
The exceptional energy efficiency of BNNs results mainly from their inherently parallel operation and their spiking behavior with associated event driven computation. Using biological tissue allows for highly efficient computing architectures. This might lead to further breakthroughs in AI and associated fields \cite{basu_organoid_2024}, despite using less electricity than current systems. 

In addition to these advantages brain organoids for use in computing could be very economical. Current costs for generating a brain organoid containing roughly $10^5$ (the number of connections is highly variable and can not specified here) cells can be as low as €0.36, with the maintenance cost for one year being around €0.66 per organoid \cite{govindan_mass_2021}. 

Nevertheless, there is no consensus on the potential costs of mature brain organoid-based computing, as there currently is no broad real-world implementation. In fact, due to the advanced and multidisciplinary technology and know-how required for such a fully-fledged system, some researchers argue that this technology may be significantly more costly when increasing its complexity \cite{basu_organoid_2024}.
\subsubsection{Mortal Computing}
Brain organoids can also be used to create self-organizing systems whose hardware and software are inseparable and unique to the system, known as \emph{mortal computers} \cite{ororbia_mortal_2023} (this might also be true for analog electronic implementation of such networks due to unavoidable parameter variations inherent to a particular implementation). These proposed systems rely on the self-organization, adaption, and homeostasis of the chosen substrate and low-level processing \cite{ororbia_mortal_2023, hinton_forward-forward_2022}. The concept of mortal computing is based on the notion of embodiment, where a system's computational power arises from the complex interplay of its physical structure and processing capabilities, such as in the case of brain organoids.

Such systems have been proposed to be highly efficient \cite{hinton_forward-forward_2022} and adaptive, \cite{ororbia_mortal_2023}, characteristics that have also been used to describe brain organoids when used for computing \cite{smirnova_organoid_2023}. Mortal computers have been proposed to be a pathway to the creation of AGI and towards advances in neuromorphic systems and computational models \cite{ororbia_mortal_2023}. With their plasticity and parallel processing, brain organoids offer a potential substrate for such systems, enabling more adaptive, efficient, and evolving computational structures.

\subsection{Challenges}
\subsubsection{Lifespan}
The biggest challenges of using brain organoids for computational purposes lie in their biological nature. First of all, brain organoids have a limited lifespan ranging from around 100 days \cite{jordan_open_2024} to 15 months and more, depending on the method of maintenance and intended application \cite{govindan_mass_2021}. 

Due to the lack of circulatory systems and the dependence on a neuronal medium to sustain them, the tissue eventually undergoes necrosis and stops developing \cite{smirnova_promise_2024}. Current brain organoid lifespans represent only 25 percent of the average life expectancy of modern supercomputers at maximum \cite{rojas_analyzing_2019}. Consequently, computing systems using brain organoids as their processing base would have a significantly shorter operational lifespan compared to traditional computers. Solving this problem is one of the most important tasks in this field \cite{sun_generation_2022}. 

Additionally, the organoids are sensitive to environmental changes and experience cell stress and necrosis even under optimal conditions, thus complicating research and application considerably \cite{zabolocki_brainphys_2020,kim_application_2023}. 
\subsubsection{Reproducibility}
Brain organoids grown from stem cells exhibit significant variability in cellular composition and structure due to their very nature and various (uncontrollable) developmental factors. Even under controlled conditions, organoids differentiate spontaneously, reducing model comparability and leading to errors, uncertain outcomes -- all significant challenges \cite{qian_brain_2019}. Their sensitivity to factors such as cell-culture conditions, generation methods, sustaining ingredients, etc., further compounds these issues \cite{poli_experimental_2019,qian_brain_2019}. This unpredictability undermines practical applications, as unreliable systems are unsuitable for critical tasks requiring accuracy \cite{mittal_survey_2016}.

In addition to these uncertainties the spikes emitted and transmitted by the neurons induce noise, further reducing the reproducibility of results. Moreover, neurons can fire spontaneously as part of their normal behavior \cite{sharf_functional_2022}. However, these spontaneous spikes typically are not a problem \cite{bakkum_spatio-temporal_2008,smirnova_organoid_2023} and can even be used as a baseline for activity in experiments \cite{cai_brain_2023}. 

Finally, the issue of reproducibility can slow down future progress and research in this field. Brain organoids' inherent spontaneity proves to be a difficult challenge to overcome in research and future of uses in a computational context.

\subsubsection{Processing Power}
Although neurons are comparatively slow and unreliable processing units, they benefit from large numbers of them being interconnected densely  \cite{ruaro_toward_2005,luczak_neurons_2022}. However, current brain organoids are still quite limited in the number of neurons available and their respective interconnectivity.

Though fully developed human brains are comparable in raw processing power to modern supercomputers, current brain organoids are only comparable to early embryonic brains with up to $\approx 10^5$ cells, whereas a fully developed human brain contains around $10^{11}$ cells with about $10^{14}$ synapses (connections) \cite{smirnova_organoid_2023}. 

Generally, higher intelligence and cognition correlates with the complexity of the brain structures \cite{goriounova_large_2018}. Accordingly, human brain organoids are currently severely limited in their capacity to replicate advanced processing as found in human brains. This will continue at least in the near future due to the absence of a circulatory system, resulting in oxygen deprivation, stress, limited maturation, and thus reduced lifespan \cite{kim_application_2023}, massively impeding their processing potential.
\subsubsection{Interfacing}
Current brain organoid computing experiments mainly rely on observing changes in the pattern of elicited spikes \cite{jordan_open_2024}. Typically, predefined regions of activity are read out and mapped according to predetermined rule sets \cite{bakkum_spatio-temporal_2008, kagan_vitro_2022}. Thus, the readout resolution is still quite limited and requires further research to make better use of brain organoids and their capabilities. The same holds true for applying input data to an organoid.

Most current experiments rely on reinforcing organoid activity with external electrical stimuli emulating rewards or punishments to train them to produce desired neural activity \cite{bakkum_spatio-temporal_2008}. However, there is no general method of reliably performing input-output operations with brain organoids \cite{jordan_open_2024}.

Current reinforcement training involves providing the brain organoids with predictable stimuli\footnote{In this case 100Hz for 100ms.} , typically with short, specific high-frequency signals, to targeted cell regions for positive feedback, and unpredictable stimuli\footnote{In this case 5Hz for 4s.}, characterized by "random" regions being stimulated with longer, low-frequency signals, for negative feedback\cite{kagan_vitro_2022}. 

In addition to this, random spikes elicited by neurons generate noise, which can result in unreliable and variable results \cite{ruaro_toward_2005}, further contributing to the issue of interpreting neural activity and interfacing with the organoids. 
There is early interfacing software, which aims to solve these issues and provide the capabilities of performing described experiments. There is early software that aims to simplify interfacing with brain organoids, for example enabling the brain organoids to control external agents \cite{tessadori_closed-loop_2015} and to be addressed remotely over the internet \cite{jordan_open_2024}.

Nonetheless, these software systems still face the same issues described above of lacking the capability to interpret complex neural activity and translating it into actionable data or addressing organoids with specific instructions.   
Through contemporary interfacing methods brain organoids can currently be trained to perform simple tasks, however, interfacing and addressing techniques are in their infancy, hindering the possibility of complex tasks. 
\subsubsection{Ethics}
The foremost ethical concern regarding experimentation on human brain organoids is the possibility of them being conscious. 
Teams of researchers have conducted analyses of experiments and research data to try to determine if current brain organoids display signs of higher-order thinking. Current research states that these organoids do not possess the complexity to allow for higher-order brain function, associated with consciousness \cite{boyd_moral_2024}. Researchers also argue that the ability to learn and complex activity signatures by themselves is not a sign of consciousness and that without a corporeal state, consciousness cannot arise \cite{croxford_case_2024}.  
However, experiments showcasing brain organoids and their plasticity, characterize them as possessing a basic form of sentience \cite{kagan_vitro_2022}. This does not constitute consciousness, as the concept remains undefined with several competing theoretical models.
Currently, the only widely acknowledged consciousness theory that allows such brain organoids to be classified as being some level of conscious, is the \emph{integrated information theory} (\emph{IIT}) \cite{montoya_what_2023}. It states that any system capable of integrating information and learning independently, is at least somewhat conscious \cite{sattin_theoretical_2021}. 

The experiments mentioned before demonstrate this learning capability of brain organoids \cite{kagan_vitro_2022, bakkum_spatio-temporal_2008, tessadori_closed-loop_2015,cai_brain_2023}. Following the IIT, brain organoids qualify as being at least somewhat conscious \cite{montoya_what_2023}. However, this assessment itself is contested as some researchers in the field argue that the IIT and its representation is basically flawed for accurately depicting possible states of consciousness in in vitro cultures \cite{kagan_neurons_2022}.
The definition of consciousness remains open, with numerous contradictory models and theories \cite{sattin_theoretical_2021,croxford_case_2024}, leaving the question unresolved whether brain organoids can feel pain or process complex emotions and sensations.

Due to ethical concerns, some countries, such as the U.S.A. and Germany,
impose regulations on research with human brain organoids, which could limit research in the future. However, in said countries, no laws prohibit research with human brain organoids. Nonetheless, guidelines for experimentation with human stem cells, must be followed and monitored \cite{pichl_ethical_2023}. If consciousness is indeed discovered in more complex brain organoids, stricter regulations could slow or even halt progress and future organoid-based computers might be entirely prohibited.
\subsubsection{Complexity}
Currently, the use of brain organoids for computing is simultaneously too complex and not complex enough for advanced tasks. The brain organoids themselves are too complex to be fully integrated and controlled with current computer systems (or fully understood), yet not complex enough to fully represent the plethora of cell types and structures found in the human brain and for more advanced processing.

Though brain organoids exhibit complex neural structures, they are obviously by far not as complex as fully developed human brains and can be compared with early stages of embryonic development \cite{qian_brain_2019} at best. This limits their current processing potential. Thus, contemporary brain organoids are vastly inferior in terms of processing power to traditional processing methods and will most likely continue to be until more developed organoids can be grown.

In addition to this writing, reading, and interpreting information to and from such organoids remains an additional challenge that hinders current developments \cite{cai_brain_2023}. While certain promising capabilities of BOs have been demonstrated experimentally, the full integration and understanding of these systems remain major challenges.
Thus, a better understanding of neural structures, their processing of stimuli, and the signals they generate is necessary. Current challenges of brain organoids for use in computing culminate in one word: complexity. The technology comes with various caveats, large challenges all by themselves, resulting in an incredibly complex set of problems to solve, if researchers want to fully understand and make use of brain organoids in a computing context, being the major stumbling block for current applications of brain organoids in a computational context.
\section{Potential Applications}
\subsection{Artificial Intelligence}
Biological neural networks like human brains can be seen as larger, more complex and ANNs with an extremely high degree of interconnectivity \cite{stoop_note_2021}, which formed the conceptual basis for digital systems such as \emph{ChatGPT}, \emph{Google Gemini} and others. The use of brain organoids, too, could heavily influence the field of AI in the next couple of years. In fact, human brain organoids have been used to conduct traditionally AI related tasks. 

Basic image processing \cite{ruaro_toward_2005} as well as the prediction of the behavior of dynamic systems exhibiting chaotic behavior \cite{cai_brain_2023} are typical tasks to which in vitro neurons have been applied. 

For the latter, such brain organoids required only 4 training epochs to achieve comparable predictive performance, whereas a machine learning algorithm with \emph{long short-term memory} (\emph{LSTM}) needed 50 epochs to outperform the biological system, representing a $>$90\% training time decrease \cite{cai_brain_2023}.

The demonstrated ability of human brain organoids to adapt to stimuli makes these functionally capable of solving more complex tasks typically associated with machine learning powered AI, such as speech recognition \cite{cai_brain_2023} or actively controlling systems such as video games \cite{kagan_vitro_2022}. 

The inherent capability of organoids to change their connection structure makes them -- at least in perspective -- more powerful than  traditional ANNs. 
Contemporary ANNs are larger and computationally intensive and therefore require pruning and specialization for efficient deployment \cite{he_structured_2023}. 
In contrast to this, the self-organized, redundant, and interconnected nature of brain organoids lends itself to being better suited for generalist AI systems \cite{stoop_note_2021,jeon_distinctive_2023}.
Human brain organoids could therefore be trained in shorter periods, given some major advances in interfacing technology, provide efficient AI solutions, and complete tasks in parallel due to the neuronal structure that constitutes them. Their ability to learn autonomously opens possibilities for the use of this technology for a wide range of tasks that currently rely on machine learning, with BNNs potentially surpassing the abilities of contemporary artificial systems.

Another area of application is the integration of classic ANNs with such organoids in order to complement each other's strengths and weaknesses. This has been coined as \emph{organoid intelligence} (\emph{OI}) \cite{smirnova_organoid_2023, basu_organoid_2024}.

Analyzing vast amounts of data remains a strong suit of ANNs \cite{badai_review_2020} and can be leveraged to characterize and analyze the vast amounts of data generated by brain organoids. This might be used in turn to better understand BNNs and their characteristics. Supplementing ANNs that excel at specialized tasks \cite{jeon_distinctive_2023} with more generalized intelligences through the application of brain organoids might eventually lead to significant advances in the field of AI \cite{cai_brain_2023, basu_organoid_2024}. Being able to learn from minimal sets of data, combined with the analytical strength of contemporary AI systems can enable superior predictive models, higher level decision-making and data analysis models \cite{basu_organoid_2024} (\emph{organoid intelligence}).

Potential real world applications have been proposed and demonstrated in the field of robotics \cite{polykretis_astrocyte-modulated_2020,independent_organoid_robot, hartung_organoid_2024}, AI algorithms in general \cite{cai_brain_2023, smirnova_organoid_2023}, such as those used for natural language processing, etc.
\subsection{Neuromorphic Computing Research}
Current neuromorphic processing units are limited by the available processing power with respect to the scope of the brain regions that they can currently simulate. Only partial (tiny) simulations of brain functionality \cite{cai_brain_2023} are possible today. Though brain organoids similarly do not represent fully developed brains and a fully diverse set of cells, current brain organoids are still more representative of portraying true brain functionality than any artificial means developed as of yet \cite{qian_brain_2019}. Accordingly, they can be used as inspiration for improved neuromorphic systems. The current and potential benefits of brain organoids can be leveraged for progress on more efficient, adaptable, and biologically representative neuromorphic processing units, see \cite{thakur_large-scale_2018}.

Contemporary research efforts in the area of neuromorphic processors almost exclusively stick to digital computer architectures, thus limiting the plasticity of the neuromorphic architectures under consideration \cite{schuman_opportunities_2022}. However, there are several recent efforts to provide such systems with new substrates that can better mimic the plasticity of BNNs, with the use of memristors being one of the more widespread approaches \cite{zhong_dynamic_2021,schuman_opportunities_2022}. 
The ability of memristive devices to change their conductance according to basically the time integral over the current flowing through the device (there are devices exhibiting thresholds to distinguish between writing and reading) and store this state \cite{zhong_dynamic_2021}, makes them highly interesting for research on neuromorphic systems. 
Memristors are currently used in tasks requiring autonomous learning and it has been demonstrated that on-chip/on-the-edge learning is a viable area of application. One particular example showed memristor based integrated circuits which learned to control a remote controlled car with input provided through a camera \cite{zhang_edge_2023}. Other examples include the recognition of written and spoken digits or words as well as the solution of dynamic systems \cite{hu_memristorbased_2018,zhong_dynamic_2021}. 

Integrated circuits with memristive devices, therefore, have successfully demonstrated the capability to emulate learning by plasticity of the underlying substrate and adapting to tasks similarly to experiments using in vitro neural cultures. Nonetheless, there are still several characteristics of biological tissue that have yet to be understood, integrated, and replicated in memristor based neuromorphic devices.

Another key characteristic of biological systems is that of cell diversity. Contemporary neuromorphic architectures focus mainly on emulating neurons in a very simplified way, neglecting the computational functions provided by other cell types critical to cognition \cite{schuman_opportunities_2022,cai_brain_2023}. 
For instance the actual behavior of dendrites is crucial in the processing of information within biological neuron, however, they are modeled in a simplified way, not fully emulating their actual function \cite{acharya_dendritic_2022}. They have been shown to be paramount in enabling continued learning and long-term plasticity \cite{acharya_dendritic_2022}. Employing better models could possibly alleviate the issue of catastrophic forgetting in ANNs, a problem for which currently no optimal solution exists \cite{jedlicka_contributions_2022}.

In addition to this, quite a lot of different cell types are completely missing from current neuromorphic systems, such as astrocytes, which play important roles in network plasticity and regulating the spiking behavior of neurons \cite{poli_experimental_2019}. Some experiments aim to simulate and integrate such missing cell types, though research in this direction is still in its infancy \cite{poli_experimental_2019}. The plasticity of BNNs stems -- at least partially -- from these different cells and the inherent prediction of possible future stimuli \cite{luczak_neurons_2022,basu_organoid_2024}. In contrast to these contemporary digital approaches, brain organoids already contain these cells, making them more biologically representative than contemporary neuromorphic systems \cite{qian_brain_2019,kagan_vitro_2022,jordan_open_2024}.

Given the limitations of current neuromorphic systems in replicating biological complexities, their adaptability to uncertainty and real-world scenarios is significantly hampered. This is partly due to the absence of a comprehensive biological representation of neural structures \cite{cai_brain_2023}.
Furthermore, while spiking associated with BNNs enhances their efficiency \cite{schuman_opportunities_2022, sharf_functional_2022}, most of neuromorphic systems of today are not of the spiking type \cite{hu_memristorbased_2018,zhong_dynamic_2021,zhang_edge_2023}, limiting their potential efficiency, although there are exceptions \cite{poli_experimental_2019,schuman_opportunities_2022}.

Current neuromorphic systems are unable to reach the levels of efficiency, plasticity, and resilience to novel unexpected inputs of BNNs. Further research and experiments on brain organoids and their application to tasks and a better understanding of the architecture underlying neural processes could advance research on substrates that can better represent biological realities. Understanding and improving  brain organoid computing technologies may serve as a vital component towards better and fully artificial neuromorphic systems.
\section{Key Research}
Maybe the most important area of research is the question of how to provide brain organoids with more abstract information and how to extract and interpret their responses. Solving these problems will advance their application in advanced AI use cases in general \cite{jordan_open_2024}. Consequently, research must focus on advancing MEA technology. This is crucial to more finely discern spikes from the organoids, in addition to allowing more precise addressing of their different cell structures. Current MEAs are severely limited with respect to the amount of addressable cells \cite{cai_brain_2023}. One proposed solution is MEAs that fully envelop the organoid \cite{martinelli_e-flower_2024, acha_neuromodulation_2025}. 

Another factor that necessitates deeper research is that of lacking reproducibility between organoids. Further methods such as those described by researchers \cite{govindan_mass_2021} are essential for this field.
Moreover, to enable longer testing and use of the organoids, research must find ways to minimize necrosis of brain organoid tissue. Possible paths to increased lifespans include improving the medium the organoids are immersed in. Alternatively, there have been experiments and proposals of fusing brain organoids with blood vessel organoids \cite{koo_past_2019, sun_generation_2022,liu_unleashing_2024}. Further implementation of such approaches might lead to longer lifetimes and thus more prolonged and complex studies involving more mature organoids. Such an advancement has the potential to increase the processing power of the brain organoids and enable more complex applications due to increased numbers of cells forming more complex structures \cite{sun_generation_2022}.

Further integration of AI tools in brain organoid experiments could help in characterizing and analyzing differences between organoids to cluster them according to behavior patterns and identify comparable groups \cite{shi_organoid_2024}. Broader research into this integration of brain organoid technology with traditional ANNs might also facilitate making use of deeper capabilities of brain organoids for computing.

All in all, research on classic neuromorphic systems and brain organoid computing should ideally work in tandem, with insights and discoveries of the function of BNNs guiding research and development of advanced neuromorphic systems. Though neuromorphic systems are based on brain functionality, research overlap between these deeply related fields remains limited as of today \cite{basu_organoid_2024}.

Early integration of ethical and legal standards is also imperative to ensure continued and unhampered research in this field. This includes exploring unresolved ethical and legal consequences of the technology. More research is necessary to determine the possible sentience and consciousness of these organoids to ensure ethical conduct of experiments. Consequently, a deeper understanding of what constitutes and distinguishes human cognition must be pursued, alongside experiments to assess the sentience of brain organoids. 

Though researchers have compared models of consciousness and evaluated the sentience of brain organoids \cite{montoya_what_2023,croxford_case_2024}, no research regarding the opinions of wider stakeholders, such as the public, policymakers, or industry leaders has been conducted. Findings on this topic can influence future ethical conduct and ensure that further progress in this field takes public and regulatory opinion into account, minimizing the risk of adverse repercussions.

Platforms like FinalSpark's Neuroplatform are designed to host and facilitate large-scale computing experiments using brain organoids offering remote access to BNNs with associated MEAs and help future research in this area \cite{jordan_open_2024}. This remotely accessible platform presents a novel opportunity for researchers worldwide to experiment with a standardized set of organoids and is already being used by several groups of researchers and universities. FinalSpark is also working on new applications of the technology, such as a demonstration of a brain organoid controlling a virtual butterfly in a 3D environment accessible to the public \cite{burger_lab-grown_2024, finalspark_lab-grown_2024}. While not a holistic solution, this platform facilitates comprehensive studies on brain organoids, their intrinsic computing mechanisms, effects of adjusting environmental conditions, all the while collecting data that can help researchers with further efforts. 

Further efforts are being conducted at Johns Hopkins University in its SURPASS program. This research is directed at several avenues of brain organoids and their computing capabilities, such as training organoids to play video games over the internet and control real-world robots \cite{hartung_organoid_2024}, with one long-term goal being the development of organoids consisting of up to one billion cells to better represent mature BNNs \cite{johns_hopkins_university_first_2024}.
\section{Discussion}
Brain organoids offer unique advantages over classic digital approaches to neural networks due to their biological nature, namely plasticity, fully parallel processing, as well as anticipatory capabilities. These features make them ideal for tasks requiring flexibility, frequent and long-term adaptation, or handling of incomplete datasets. These features facilitate applications such as natural language processing, robotics, and energy-efficient multitasking systems. Practical use cases include generalized machine learning, pattern recognition, and simulations of dynamic systems, with a focus on leveraging their intrinsic long-term learning ability and low power consumption. 

However, brain organoids face significant challenges, including limited lifetime, low reproducibility, complex interfacing, and ethical concerns. As of now, they are unsuitable for vast data processing or tasks needing extensive computational power. Inconsistent development and signaling among organoids further hinder standardization and debugging. Ethical oversight and compliance with regulations are also crucial to sustaining progress in this field. Applications should prioritize tasks aligned with the technology's strengths while minimizing risks and addressing these limitations.

Human brain organoids have been shown to be capable of highly efficient, parallelized computation, with extrapolation capacities that ANNs lack, all while current experiments are rather low-cost. Nevertheless, these capabilities are hindered by above-mentioned challenges. The development and application of human brain organoids is still in its formative stages, with current in vitro cultures still being rather limited. These issues are compounded by challenges of reliably reproducing results, due to the heterogeneous nature of brain organoids. Furthermore, ethical and legal implications of experimenting on and using human brain cells as a tool for computations necessitate close monitoring and work in accordance with current regulations for further research to be possible. 

Despite the potential and opportunities offered by this emerging technology, due to the current limitations and shortcomings of human brain organoids in computing, their application in real-world problems is still far in the future. However, basic capabilities and proofs of concept allowing for such applications have been demonstrated in recent works. For now, progressing the emulation of brain functionality and developing more advanced brain organoids to learn from them may present itself as the most practical approach in advancing research and refining the technology.

In the end, if the biological complexities of these systems are solved, human brain organoid computing could become a new frontier for multidisciplinary computing technologies. For the time being, however, researchers must focus on fundamental research to build on existing findings and overcome existing challenges to move forward to real-world applications. The full potential of brain organoid technology in computing is yet to be realized, offering a range of possibilities to be explored in future research.  
\printbibliography
\vspace{-1cm}
\begin{IEEEbiography}[{\includegraphics[width=1in,height=1.25in,clip,keepaspectratio]{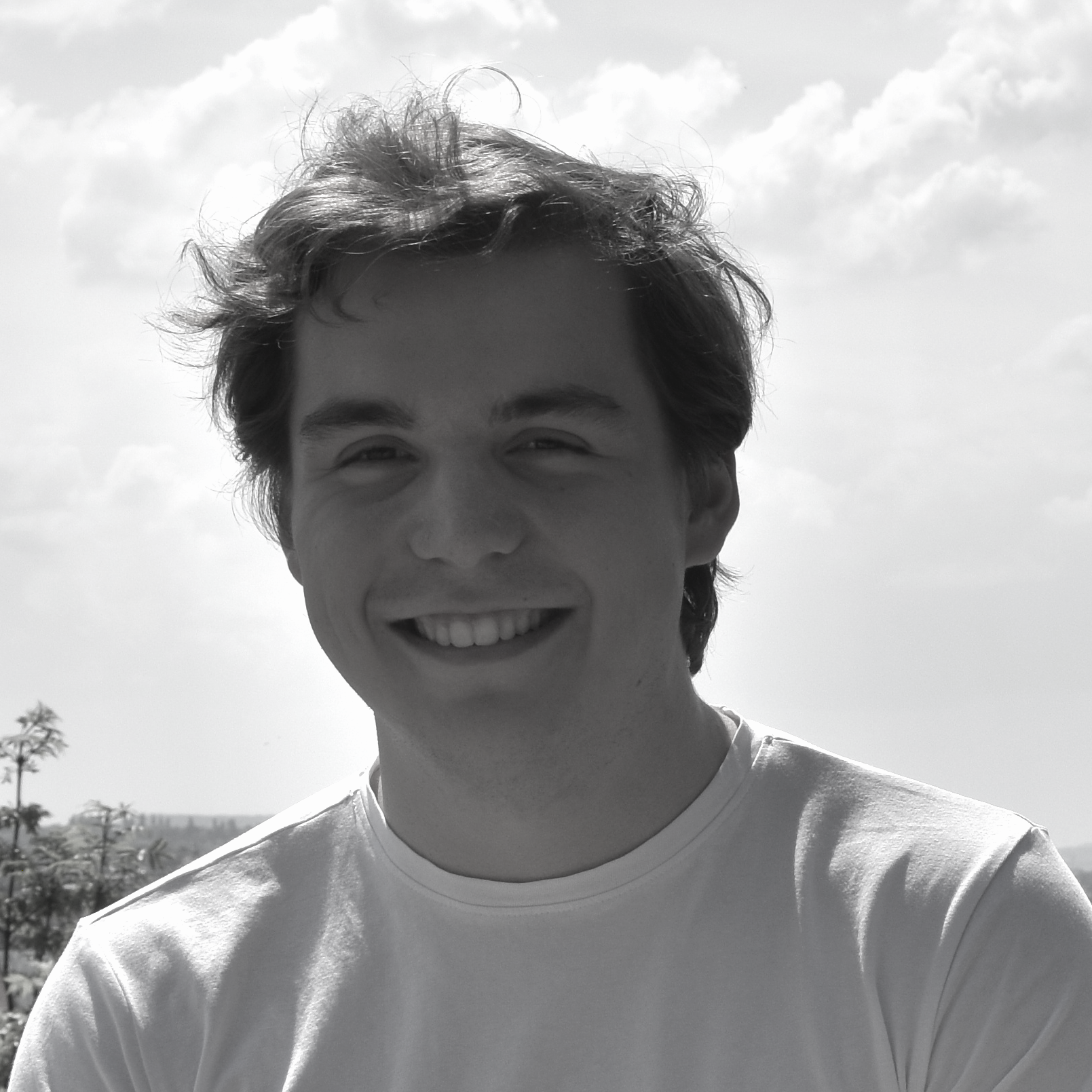}}]{Yannic Talavera}
was born in Worms, Germany in 2002. He completed his bachelor's degree in Business Informatics at the FOM University of Applied Sciences in 2024. He is currently pursuing his Erasmus Mundus Joint Master's Degree in Artificial Intelligence (EMAI). His academic interests include unconventional computing approaches, cognitive science, and their interdisciplinary applications.
\end{IEEEbiography}

\vspace{-1cm}
\begin{IEEEbiography}[{\includegraphics[width=1in,height=1.25in,clip,keepaspectratio]{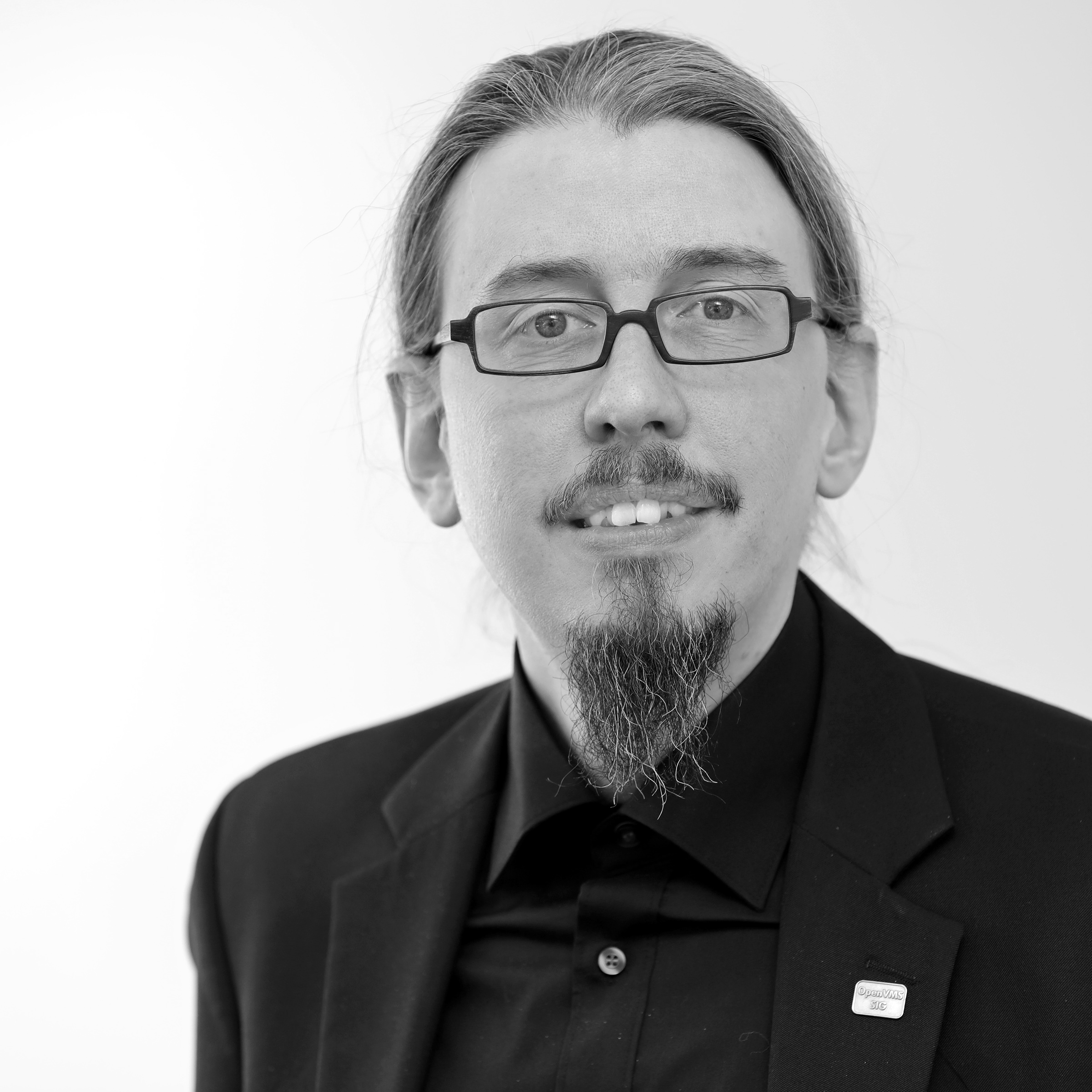}}]{Bernd Ulmann}
was born in Neu-Ulm, Germany in 1970, and received his diploma in mathematics from the Johannes Gutenberg-Universit\"at Mainz, Germany, in 1996. He received his Ph.D. from the Universit\"at Hamburg, Germany, in 2009. Since 2010, he has been professor for business informatics at the FOM University of Applied Sciences, Frankfurt/Main, Germany. His main interests are analog and hybrid computing, the simulation of dynamic systems, and operator methods. He has authored several books on analog and hybrid computing.
\end{IEEEbiography}
\end{document}